# Traceably Calibrated Scanning Hall Probe Microscopy at Room Temperature

Manuela Gerken[1*], Aurélie Solignac[2], Davood Momeni Pakdehi[1], Alessandra Manzin[3], Thomas Weimann[1], Klaus Pierz[1], Sibylle Sievers[1], and Hans Werner Schumacher[1]

[1]Physikalisch-Technische Bundesanstalt (PTB), Bundesallee 100, D-38116 Braunschweig, Germany
[2]SPEC, CEA, CNRS, Université Paris-Saclay, CEA Saclay, F-91191 Gif sur Yvette Cedex, France
[3]Istituto Nazionale di Ricerca Metrologica, Torino, I-10135, Italy

## Abstract

Fabrication, characterization and comparison of gold and graphene micro- and nano-size Hall sensors for room temperature scanning magnetic field microscopy applications is presented. The Hall sensors with active areas from 5 µm down to 50 nm were fabricated by electron-beam lithography. The calibration of the Hall sensors in an external magnetic field revealed a sensitivity of 3.2 mV/(AT) ± 0.3 % for gold and 1615 V/(AT) ± 0.5 % for graphene at room temperature. The gold sensors were fabricated on silicon nitride cantilever chips suitable for integration into commercial scanning probe microscopes, allowing scanning Hall microscopy (SHM) under ambient conditions and controlled sensor-sample distance. The height dependent stray field distribution of a magnetic scale was characterized using a 5 µm gold Hall sensor. The uncertainty of the entire Hall sensor based scanning and data acquisition process was analyzed allowing traceably calibrated SHM measurements. The measurement results show good agreement with numerical simulations within the uncertainty budget.

## I. Introduction

High resolution quantitative magnetic field measurements at the micrometer scale are increasingly important for research and development in areas like magnetic sensors and magnetic positioning. However, the characterization of microscale magnetic structures entails new challenges for magnetic stray field measurement techniques, since the generated magnetic stray fields locally change their direction on the nanometer range, and the field amplitude decreases rapidly with an increasing distance to the sample surface. Thus, suitable magnetic sensors not only need to be small to avoid averaging over different stray field directions but also must be precisely positioned close to the sample.

Here, micro- and nano-scale Hall sensors for traceable scanning Hall probe microscopy (SHPM) are characterized, and one technique to integrate them into a commercial atomic force microscope (AFM) is presented. AFM based SHPM[1–3] (AFM-SHPM) has certain advantages compared to other magnetic imaging techniques. In comparison to magnetic force microscopy[4,5], AFM-SHPM can be considered as non-invasive due to the use of non-magnetic materials and a neglectable magnetic field produced by the supply current. A measurement directly generates quantitative results albeit with reduced spatial resolution. It is applicable to a broader field range than magneto-optical indicator film techniques[6,7], which are limited by the saturation field of the sensor film. Unlike magneto-resistive sensors[8,9], Hall probes show excellent linearity without hysteresis and measure a well-defined field component perpendicular to the sensor plane. Furthermore, SHPM enables a low sample-probe distance and therefore a spatial resolution limited by the sensor size only. A three-dimensional mapping of the stray field can be performed by repeating the two-dimensional plane scans at defined heights[10] and first commercial systems employing sub-micron Hall sensors for room temperature (RT) measurements are available[11,12]. Concerning the choice of the Hall sensor material different material classes have been considered: Standard Hall sensor devices are typically based on semiconductor technology. They show an outstanding performance at low temperatures[13], and micron-sized sensors are also working at room temperature[1,14,15]. In contrast, nano-scale semiconductor Hall sensors show a weak signal to noise (S/N) ratio at RT. Besides the semi-metal bismuth[3], graphene[2,16–19] has the potential to bridge this gap. The semi-metallic graphene can be prepared with a low carrier density at RT resulting in a high Hall coefficient. It could also enable the application of Hall sensors at high temperatures[16], in contrast to semiconductor sensing with a limited range of operational temperatures. Moreover, a one atom thin active layer prevents field averaging perpendicular to the sensor area allowing high spatial resolution in the corresponding direction. Another option is the use of metals as Hall sensor materials. Here, in particular, gold is favorable due to its good electrical properties and its chemical stability. It offers a very stable carrier density which is important for calibrated Hall measurements. A stable carrier density leads to a stable Hall coefficient and thus to a low uncertainty of the Hall sensitivity. Its carrier density is widely insensitive to surface contaminations as well as to the temperature, in contrast to semiconductors. Gold is simple to manage in the fabrication process. The fabrication is cheaper and less time consuming than for semiconductor-based Hall sensors. The substrate material can be chosen flexible due to the simple deposition process of gold. This enables the fabrication of gold Hall sensors on cantilever tips and consequently the performance of AFM-SHPM. Strain induced in the sensor due to the cantilever oscillation has no influence on the electronic structure and thus on the Hall signal whereas this cannot be neglected for semiconductors. Additionally, it was reported that gold Hall sensors with sizes below 500 nm have a better noise figure at RT than sensors based on two-dimensional electron gases[20].

---

[*] E-Mail address of the corresponding autor: Manuela.gerken@ptb.de

In this work, the fabrication and investigation of gold- and graphene-based micro- and nano-Hall sensors with respect to SHPM applications is presented. The sensor sensitivity and stability are characterized and discussed. Gold Hall sensors are fabricated on silicon nitride (SiN) cantilever chips suitable for AFM-SHPM. After traceable calibration, they allow quantitative stray field measurements of magnetic scales with few microns resolution. The uncertainty budget of the measurements is discussed, and the setup is validated by comparing the measurement results to stray field and Hall voltage signal calculations.

## II. Fabrication of Hall sensors

The Hall sensor cross-structures with active square areas from 5 x 5 µm² down to 50 x 50 nm² were fabricated using electron-beam lithography (EBL). The substrate of the gold sensors consists of a silicon wafer covered on both sides with a 1 µm thick SiN layer deposited by low-pressure chemical vapor deposition. The inset of Figure 1a shows a gold cross-shaped Hall sensor that is produced through two lithography steps. First, the Hall cross defining the sensor's active area is structured, after electron beam deposition of the metals, by EBL and lift-off through a PMMA resist. The active sensor material consists of a 5 nm titanium adhesion layer followed by a 30 nm gold layer. In the second step, an additional 50 nm gold layer is deposited to support the outer extended contact regions. After fabrication of the Hall sensors, cantilever chips with Hall sensors on the tips of the cantilevers, as depicted in Figure 1, are fabricated out of the SiN wafer by optical lithography and etching. The shape of the cantilever is defined by an aluminum (Al) mask, resistant to the SiN final reactive ion etching (RIE). Windows are opened in the SiN on the backside by RIE allowing the wet etching of silicon under the cantilever region with potassium hydroxide (KOH). The cantilevers now stand up on windows of 1 µm thick SiN and are protected by an Al mask. Cantilevers are then released by performing the final top RIE etching of the SiN, and the Al mask is dissolved in a photoresist developer basic solution. By careful alignment of the Hall sensor and cantilever fabrication processes, a Hall sensor positioning close to the cantilever tip is possible. The geometrical dimensions are visualized in Figure 1b and c. For the 5 µm sensors a minimal distance between the sensor center and triangular cantilever tip of 20 µm and for the 50 nm ones a distance of 2.4 µm can be achieved.

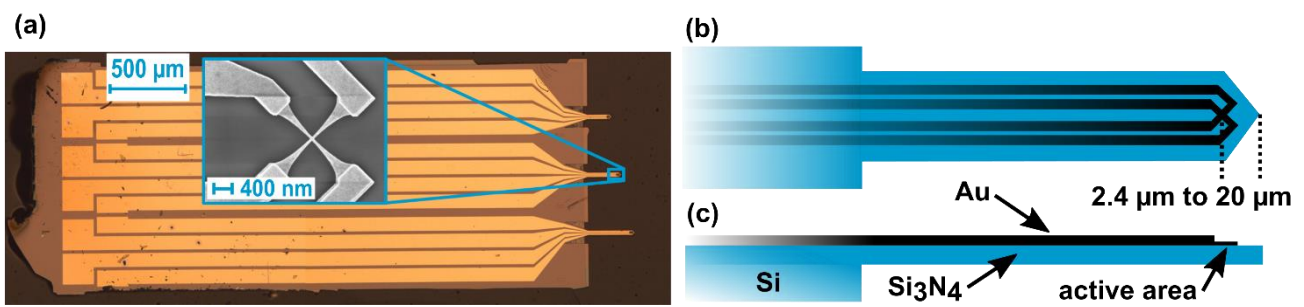

*Figure 1.* (a) Microscopy image of a 3.4 mm x 1.5 mm cantilever chip with three gold Hall sensors for integration into commercial scanning probe microscopy systems. The inset shows a scanning electron microscopy image of a 50 nm gold Hall cross. (b) Top view sketch of the Hall sensor on the cantilever. The distance between the active area and the triangular tip of the cantilever depends on the size of the sensor, so that the smaller sensors can be positioned closer to the tip. c) Side view of the cantilever. Dimensions are exaggerated for the sake of visibility (the cantilever is 1 µm thick).

The graphene samples were grown on silicon carbide (SiC) (0001) substrates with a size of 5 x 10 mm$^2$ using a so-called polymer-assisted sublimation growth technique[21–23]. The high morphological and electronic homogeneity of the graphene samples utilizes scalable realization of Hall sensors on true two-dimensional carbon sheets without bilayer inclusions. The graphene Hall sensors were patterned with EBL and AC plasma-etching through a resist mask. The fabrication of graphene Hall cross-structures requires four steps. Initially, the electrical contacts are defined by depositing Ti/Au (10/30 nm) layers in a lift-off process. In the next step, the monolayer graphene is structured by plasma etching through an EBL resist mask, leading to well-defined geometry and sizes of the small active area as shown in Figure 2. The finger-shaped contact structure produces a long contact line between graphene and gold in the plane which enables an efficient charge transfer and reduces the contact resistance[24–26]. Finally, the outer contact areas and leads are defined using a 50 nm gold layer. To avoid the environmental influences, especially the rapid change of the carrier density by surface absorption on the graphene, the graphene sample was encapsulated with 50 nm co-polymer. Note that within this work the graphene sensors were only fabricated on SiC wafer dies. The fabrication of graphene sensors on cantilevers will be subject of future studies.

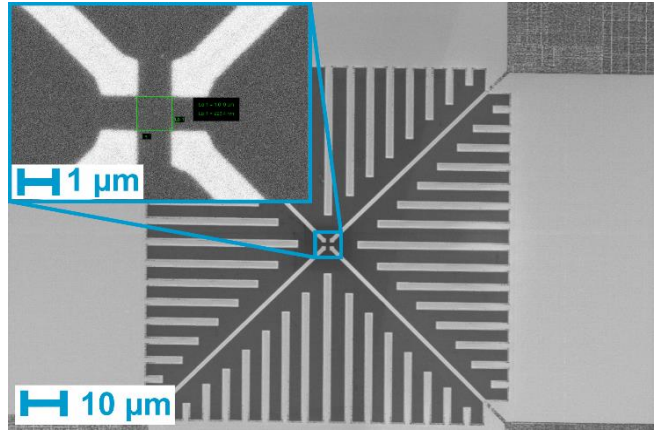

*Figure 2. Scanning electron microscopy image of a 1 µm graphene Hall cross and the contact region before deposition of the final gold layer. The inset shows a magnification of the cross area.*

## III. Characterization of the Hall sensors

For the Hall sensor calibration, an electromagnet driven by a Kepco power supply with a pole shoe diameter of 92 mm was used to provide a spatially homogeneous magnetic flux density up to 450 mT at a pole shoe distance of 18.5 mm. The operation current for the Hall sensors was generated by a Keithley source meter 2400. The Hall voltage was measured with Keithley Nanovoltmeter 2182A. During the Hall sensor calibration, the magnetic flux density was simultaneously measured with a traceably calibrated commercial Hall probe FH55 from Magnet-Physik Dr. Steingroever GmbH. As a consequence, the Hall sensor calibration is traceable to the SI units when considering and listing all uncertainty contributions, as presented in section VI.

The typical output from the characterization of graphene and gold sensors is presented in Figure 3. The Hall voltage was measured as a function of the magnetic flux density $B$ and corrected for the offset. Both sensors show a linear dependence of the Hall voltage $V_{\text{Hall}}$ on $B$ as expected from $V_{\text{Hall}} = I \times B/(n \times e \times t)$, where $I$ is the supply current, $n$ is the electron density, $e$ is the electron charge and $t$ is the thickness of the active layer. For the 5 µm gold Hall sensor operated at 10 mA, the output is in the µV range for $B$ between -150 mT and 150 mT. This leads to a sensitivity of 3.2 mV/(AT) ± 0.3 %. For the same field range, the Hall voltage of the 500 nm graphene sensor is in the mV range using an operating current of 50 µA. Fitting the data reveals a sensitivity of 1615 V/(AT) ± 0.5 %. The sensitivity of the graphene sensor is six orders of magnitude higher due to the lower carrier density of graphene in comparison to gold. Similar results were observed in the measurements on several other 5 µm large Hall sensors. The mean sensitivity of all gold sensors is 3.1 mV/(AT) with a maximum deviation of 0.2 mV/(AT) within the sensors in this study.

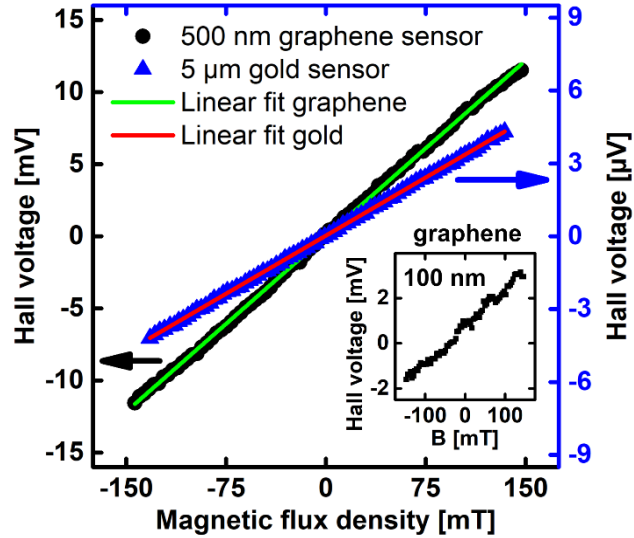

*Figure 3.* Typical calibration curves of gold and graphene Hall sensors. For both measurements, the offset was subtracted from the Hall voltage. The 500 nm graphene Hall sensor was operated at 50 µA and exhibited a sensitivity of 1615 V/(AT) ± 0.5 %. A supply current of 10 mA was used for the 5 µm gold sensor. The sensitivity extracted from the fitted result is 3.2 mV/(AT) ± 0.3 %. The inset shows the calibration result of a 100 nm graphene sensor operated at 10 µA. The sensitivity was determined to 1649 V/(AT) ± 1.17 %.

Moreover, the time stability as well as fabrication reproducibility of the sensors were investigated. To this end, the sensors were frequently characterized within one year and compared with nominally identical sensors from different batches. For the gold sensors, the long-term stability was very high with a deviation over time below 0.6 %. Graphene sensors showed sensitivity deviations of up to 9.3 % from one day to another. This might be caused by charges attaching and detaching on the co-polymer which act like a gate and thus influence the carrier density in the graphene. Also, the overall variation in sensitivity was larger for graphene sensors ranging from 500 V/(AT) to 1700 V/(AT) depending on the carrier density in the respective graphene material and actual sensor. Based on the Hall voltage deviation of measured data points from the expected value given by the linear characterization fit, a typical resolution of 2 mT for gold sensors and 0.45 mT for the graphene sensors is calculated. This resolution includes, besides the sensor properties, also influences and noise contributions from devices and cables in the circuitry. Noise measurements revealed a detectivity of $60~\mu T/\sqrt{Hz}$ at 1 Hz for a graphene Hall cross. The property data are summarized in Table I. Due to the small resistance of the gold Hall sensors, it was not possible to measure their noise characteristics. This also means that the noise properties of the electronics have a larger influence on the S/N than the sensor itself. With the described measurement equipment, it was possible to calibrate gold sensors with active areas down to 1 µm. For 50 nm gold sensors on cantilevers, the background noise of the setup is larger than the expected Hall voltage of 1.5 nV per 10 mT at the operating current of 50 µA. Because of higher sensitivity and thus larger Hall voltage, graphene sensors with a size of 100 nm still show an overall linear dependence on the applied magnetic flux density, as shown in the inset of Figure 3. The resolution is decreased to 8.5 mT because of the growing impact of carrier fluctuations for smaller sensor sizes. Furthermore, the lower current supply leads to smaller Hall voltages, thereby the overall S/N is reduced. This comparison of gold- and graphene-based Hall sensors emphasizes the benefits and drawbacks of metallic and semi-metallic Hall sensors with the same small active areas and under consideration of the uncertainty budget.

*Table I. Properties of 5 µm graphene and gold Hall sensors.*

| Sensor material | Sensitivity | Resolution | Detectivity @ 1 Hz | Long-term deviation |
|---|---|---|---|---|
| Gold | 2.9 – 3.3 mV/(AT) | 2 mT | - | 0.6 % |
| SiC/Graphene | 500 – 1700 V/(AT) | 0.45 mT | 60 µT/$\sqrt{\text{Hz}}$ | 9.3 % |

## IV. AFM based SHPM – setup and measurement

AFM-SHPM is realized by integration of the manufactured cantilever chips with gold Hall sensors into a commercial AFM (Nanoscope IIIa, Dimension 3000 scanner). The cantilevers have a typical resonance frequency of about 50 kHz and can be used in standard tapping mode operation, and thus in close contact with the sample surface. As shown in the upper part of Figure 4, the Hall sensors are positioned at the bottom side of the cantilever and close to its tip to achieve a small distance between the sensor and sample. This is significant for improving spatial resolutions in measuring nanostructures due to the fast decay of stray fields with increasing distance to the sample surface. The cantilevers are mounted at an angle of 10° given by the cantilever holder. This leads to minimal measurement heights of 400 nm and 3366 nm for the 50 nm and 5 µm sized sensors, respectively, for an ideal alignment of the Hall sensor on the cantilever. The electrical connection to the Hall sensor is realized by bond wires from the cantilever chip to a printed circuit board (PCB) that is fixed to the cantilever holder. The current source, voltmeter and PCB are connected via soldered cables. To increase the scan area up to millimeter range, additional piezo tables were added to the setup that allow scanning the sample with a fixed cantilever position. The Hall sensor was calibrated in the electromagnet before and after the AFM-SHPM measurement. As a test sample, a commercially available magnetic scale SST250HFA-04 from Sensitec was chosen. By the selection of this sample, the operation of AFM-SHPM can be demonstrated on a real and industrially relevant magnetic sample. Thin film conductors or coils, as known from MFM calibrations[27], are not able to generate suitable stray fields of about 100 mT on the micrometer scale. The commercial magnetic scale is made of a wet pressed strontium ferrite with a remanence magnetization of $M_r$ = 395 mT. The material was magnetized into alternating up and down magnetized stripes with a width of nominally 250 µm and several millimeters length.

The bottom part of Figure 4 displays the results of AFM-SHPM on the scale using a 5 µm gold sensor with a Hall sensitivity of 2.3 mV/(AT) ± 13 %, measured under an applied operating current of 1 mA. Line scans with ten repetitions each were performed at seven different measurement heights. The closest line scan to the sample was attained in the tapping mode and thus followed the sample topography at a distance of approximately 4 µm. The scans for higher distances were carried out at fixed heights ranging from 19 µm to 169 µm (with 30 µm intervals) by moving the sample with a 3-axis piezo scanning system and fixed probe position. The sample has a granular structure and thereby height variations of around 10 µm and locally tilted surface areas. Therefore, during the measurement in tapping mode, the z-piezo table was utilized to keep the sensor-sample distance in a range that is controllable by the AFM head. The measurement was performed at a scan speed of 50.5 µm/s. 400 points were measured per line with an averaging time of 20 ms. All plots in Figure 4 show the expected 500 µm periodicity of the scale. The decay of the stray field amplitude with increasing distance to the sample is clearly visible in both actual measurement values and simulation results, as presented in the next paragraph. Furthermore, cantilever and sensor appeared to be very robust, allowing the characterization of rather rough samples, as demonstrated in this study.

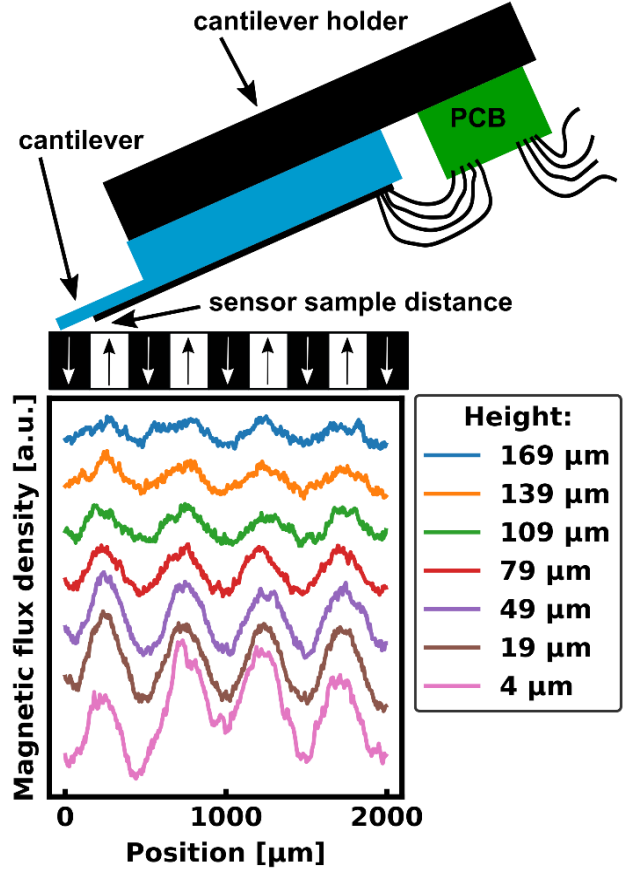

*Figure 4*. Schematic of the AFM-SHPM measurement principle with electrical connections to the sensor and the minimal sensor-sample distance. Below this, the height dependent SHPM of a magnetic scale with 250 µm pole width is shown. A 5 µm x 5 µm gold Hall sensor scanned one line of the sample 10 times for each height. The measurement at 4 µm was performed in tapping mode so that the cantilever tip was in contact with the sample surface. Data from the different measurement heights were shifted to avoid overlapping.

## V. Comparison with simulations

Here, two modeling approaches are presented to validate the measurement procedure. The first one uses a Fourier transform method to calculate the *z*-component $B_z$ of the stray field produced by the magnetic scale. Assuming that the sample is perfectly parallel to the Hall sensor surface and neglecting the 10° cantilever tilt, the Hall sensor response is mainly dominated by the perpendicular component. From now on, this will be called the perpendicular or z-component in contrast to the in-plane components lying parallel to the sample surface. The calculated field profile is then compared to the experimental one, which is derived from the measured Hall voltage as $B = (V_{Hall} \times n \times e \times t)/I$ after the application of offset corrections. In the following, the stray field simulation procedure is described. To simulate the perpendicular stray field component, the underlying sample magnetization has to be known. To this end, the magnetization is guessed from the measured Hall signal based on the following assumptions: (i) The transition between up and down magnetized poles can be found at zero transitions of the Hall signal after a performed offset and drift correction. (ii) The magnetization pattern of the scale, therefore, can be found by discrimination between areas with positive and negative Hall voltages. (iii) Areas with $+V_{\mathrm{Hall}}$ ($-V_{\mathrm{Hall}}$) have purely perpendicular and, over the thickness, homogenous magnetization of $+M_r$ ($-M_r$). To account for the pole writing process, the step-like transitions are then additionally smoothed as visualized in Figure 5a. For the stray field calculation, a transfer function (TF) approach[28,29] was pursued due to its numerical simplicity. It can be shown, that in a partial Fourier space of only the in-plane spatial components, as for the transformation from ($x,y,z$) to ($k_x,k_y,z$), the calculation of the stray field above a perpendicular magnetization distribution can be performed through a multiplication by a $M_z$ to $B_z$ TF.

$$B_z(\mathbf{k}, z) = \mu_0 M_z(\mathbf{k}) \times \frac{(1 - e^{-kd})e^{kz}}{2} \tag{1}$$

The stray field of the scale at measurement height $z$ was calculated for a magnetic layer with an assumed thickness of $d$ = 75 µm. Similarly, the impact of the finite sensor dimensions might be considered by introducing a multiplication by an appropriate sensor sensitivity TF. However, for the relatively slowly varying field of the scale with 250 µm pole width, this is expected to have a minor impact and was therefore neglected. Inverse discrete Fourier transformation was used to obtain the value of the perpendicular component plotted in Figure 5a for each SHPM measured point in real space.

A good agreement between the measured stray field and simulated data was obtained, giving evidence of the validity of the measured quantitative magnetic field distribution, as shown exemplarily in Figure 5a for the sensor-sample distance of 49 µm. The simulation confirms very well the measurement results in terms of maximal and minimal magnetic flux density as well as spatial periodicity. However, the drift and slight decrease in amplitude cannot be explained by this stray field simulation. One reason would be an unstable temperature during the measurement and thus a drift of the offset voltage. Another explanation would be an angular misalignment of the sample if it is not flat or placed perfectly horizontal on the table. Moreover, the implementation of the cantilever chip in the AFM with a 10° canting angle, resulting in a canted Hall sensor with respect to the sample surface, would lead to an asymmetric signal. As a proof, a second modelling approach was implemented, in which the Hall voltage signal due to magnetic scale scanning is numerically calculated, considering an angular misalignment of about 1° between the magnetic sample and the Hall device. The spatial distribution of the electric potential $\phi$ within the sensor is derived from the finite element solution[30] of the following equation

$$\nabla \cdot [\ddot{\sigma}(\mathbf{r}) \nabla \phi(\mathbf{r})] = 0, \tag{2}$$

where $\ddot{\sigma}(\mathbf{r})$ is the conductivity tensor

$$\ddot{\sigma}(\mathbf{r}) = \frac{\mu n e}{1 + [\mu B_\perp(\mathbf{r})]^2} \begin{bmatrix} 1 & \mu B_\perp(\mathbf{r}) \\ -\mu B_\perp(\mathbf{r}) & 1 \end{bmatrix}. \tag{3}$$

In (3) $\mu$ is the electron mobility, assumed equal to 8.7 x 10$^{-4}$ m$^2$/(Vs) from 4-point resistance measurement, $n$ = 1.92 x 10$^{21}$ m$^{-2}$ and $B_\perp$ is the orthogonal component to the sensor of the stray field from the scale below, which also includes the component of $B$ parallel to the sample surface, due to the sensor-sample relative angular orientation. The formulation is completed by the boundary conditions in correspondence of the current and voltage contacts.

The stray field from the scale, which is discretized in $N$ 10 µm size hexahedra with imposed uniform magnetization, is calculated as

$$\boldsymbol{B}(\boldsymbol{r}) = \frac{\mu_0}{4\pi} \sum_{e=1}^{N} \int_{\partial \Omega_e} \boldsymbol{M}(\boldsymbol{r}_e) \cdot \boldsymbol{n}_e \frac{(\boldsymbol{r} - \boldsymbol{r}_e)}{\|\boldsymbol{r} - \boldsymbol{r}_e\|^3} ds, \qquad (4)$$

where $\partial \Omega_e$ is the surface of the $e$-th hexahedron having normal unit vector $\boldsymbol{n}_e$ and barycenter with vector position $\boldsymbol{r}_e$[31].

The drift effect in the measured Hall voltage signal is well reconstructed by the numerical results, which also support the validity of the linear dependence of $V_{\text{Hall}}$ on $B$ for all the scanning points, due to the large width of the pole scale with respect to the Hall cross size. The agreement with experimental results is highlighted in Figure 5b for an average sensor-sample distance of 49 µm and 139 µm. The peaks reduce in amplitude during scanning, as a consequence of the increase in the sensor-sample distance.

For further verification, the behavior of the stray field with an increasing distance to the sample surface, as shown in Figure 4, was quantitatively analyzed. Therefore, the maximal measured stray field amplitudes over the poles for each measurement height were compared in Figure 6 with the values expected from simulations with the first approach. For the two largest measurement heights and corresponding two lowest expected magnetic flux densities, systematic uncertainties from evaluating the extrema have a more significant influence on the result due to an enlarged contribution of noise. However, for all measurements, the simulation result overlaps with the uncertainty squares of the data points. More details about the uncertainty range are given in the next paragraph. From these results, the validity of the quantitative AFM-SHPM method using a gold sensor is concluded.

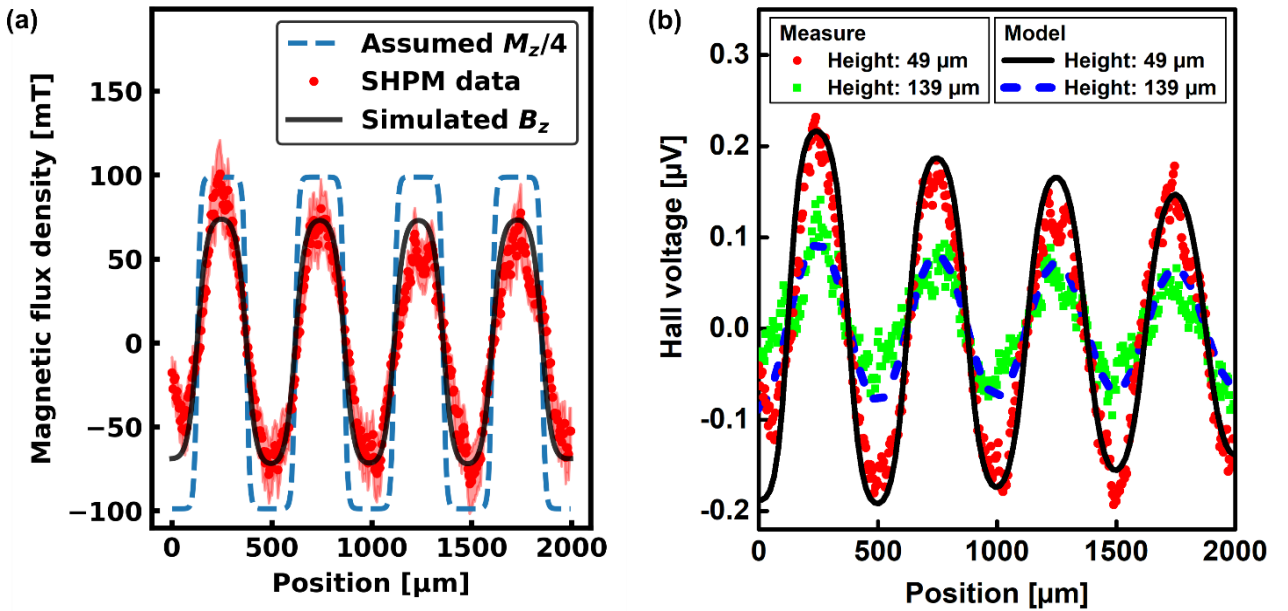

*Figure 5.* (a) Traceable SHPM data with uncertainty budget of scale SST250HFA-04 at a measurement height of 49 µm compared to simulated stray field values from the assumed magnetization distribution. (b) Comparison of experimental and calculated Hall voltage signal for measurement heights of 49 µm and 139 µm. The calculation is made with the second modelling approach, introducing an angular misalignment of about 1° between the magnetic sample and the Hall device to explain the drift during scanning.

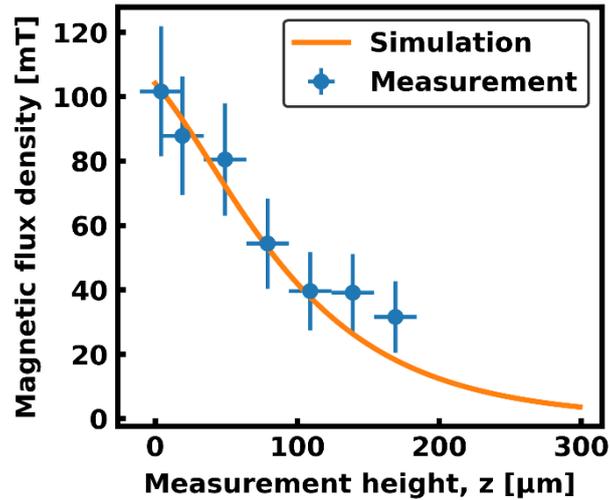

*Figure 6*. Simulated stray field decay with increasing measurement height over the middle of a pole. The data points are generated from the measurements shown in Figure 4 by averaging the absolute values of the three minima and four maxima.

## VI. Evaluation of uncertainty budget

A traceable scanning Hall sensor calibration obligatorily requires an analysis of all contributions of the entire scanning and read-out process to the uncertainty and a statement of their values. Five major contributions entering into the net measurement uncertainty are: (i) The Hall sensor itself, where its stability, sensitivity, offset, temperature dependence and noise must be considered. (ii) The Hall sensor calibration via the electromagnets magnetic field homogeneity, stability and repeatability. (iii) The Hall sensor driving and read-out electronics, including the stability of the current source and the voltmeter noise as well as thermoelectric voltages. (iv) The positioning accuracy of the scanning system. (v) Influence of the sensor on the sample, for example in terms of the magnetic stray field generated by the supply current. The multiplicity of uncertainty sources and the fact that standard uncertainty analysis is not sufficient for linear regression tasks[32], as used in the Hall sensor calibration, rule out a conventional uncertainty propagation calculation. Therefore, the uncertainties of the main contributions were analyzed separately to evaluate their impact on the measurement.

By repeated sensor calibrations and statistical analyses of the results, a calibration uncertainty for the sensor sensitivity of 13 % was found. The characterization of several gold sensors, as shown above, leads to the expectation of lower sensitivity uncertainties for the gold sensors in general. The uncertainty includes contributions from the electronic components and the applied magnetic flux density. The scanning system uncertainty is 7 mT after the correction of drifts by offset subtraction. This was evaluated from ten repeated line scans. The different measurement heights were realized by an additional z-piezo that has a positioning accuracy of 15 µm. The stray field produced by the sensor supply current at the sample surface was estimated with the help of Biot-Savart law to 0.25 mT. This has no influence on the sample characterized here and is neglectable for many other applications. Combining all contributions leads to an extended uncertainty of ± (7 mT + 13 %) for the SHPM using the amplification factor k = 2.

# VII. Conclusion

In summary, SiN based AFM cantilevers equipped with micro- and nanoscale gold Hall sensors were fabricated, which facilitate accurate traceably calibrated scanning magnetic field microscopy (AFM-SHPM) at room temperature. The measurement data were in good agreement with simulation results, which underline the reliability of the presented approach. The gold sensors exhibit a sensitivity of 3.2 mV/(AT) with high long-term stability. Also, Hall sensors out of epitaxial, zero-band gap, semi-metallic graphene (on SiC) were fabricated and studied. In contrast to the metallic gold sensors, the graphene samples show an outstanding high sensitivity of 1700 V/(AT), but low time-stability. This suggests that by proper isolation of the graphene sensors from environmental influences (e.g., using hexagonal boron nitride or aluminum dioxide) a higher performance could be achieved. This, in addition to the implementation of the graphene Hall sensor into the AFM cantilever, are the subjects of future studies. For the 5 µm gold AFM-SHPM, the uncertainty budget of the entire room temperature measurement process was analyzed and determined to be ± (7 mT + 13 %). This method enables a direct quantitative characterization of magnetic microstructures in ambient conditions with the capability of generating three-dimensional maps of the sample's out of plane stray fields within a range from mT to few T. We expect, that by developing a suitable, low-noise electronic for the 50 nm gold Hall sensors the spacial resolution can be further increased which will enable the direct traceable characterization of magnetic nanostructures at room temperature. The fabrication process is scalable thus in principle allowing high volume sensor production. Finally, the AFM-SHPM is a non-destructive and robust method for both scientific research as well as industrial applications, e.g., quality control within industrial processes.

## Acknowledgments

This work was supported by the EMPIR JRP 15SIB06 NanoMag through EU and EMPIR participating countries within EURAMET. Manuela Gerken gratefully acknowledges the support of the Braunschweig International Graduate School of Metrology B-IGSM and the DFG Research Training Group 1952 Metrology for Complex Nanosystems. We thank Felix Nording from Institut für Elektrische Messtechnik und Grundlagen der Elektrotechnik of Technische Universität Braunschweig in Germany for performing the noise measurements. Davood Momeni Pakdehi acknowledges support from the School for Contacts in Nanosystems (NTH nano).